# Deep RAN:

# A Scalable Data-driven platform to Detect Anomalies in Live Cellular Network Using Recurrent Convolutional Neural Network

1st Mohammad Rasoul Tanhatalab
RAN Performance
*MTN IranCell*
*Tehran, Iran*
mohammadrasoul.t@mtnirancell.ir

2nd Hossein Yousefi
RAN Performance
*MTN IranCell*
*Tehran, Iran*
Hossein.yous@mtnirancell.ir

3rd Hesam Mohammad Hosseini
RAN Performance
*MTN IranCell*
*Tehran, Iran*
hesam.mo@mtnirancell.ir

4th Mostafa Mofarah Bonab
RAN Performance
*MTN IranCell*
*Tehran, Iran*
mostafa.mof@mtnirancell.ir

5th Vahid Fakharian
RAN Performance
*MTN IranCell*
*Tehran, Iran*
vahid.fak@mtnirancell.ir

6th Hadis Abarghouei
RAN Performance
*MTN IranCell*
*Tehran, Iran*
hadis.ab@mtnirancell.ir

*Abstract*— In this paper, we propose a novel algorithm to detect anomaly in terms of Key Parameter Indicators (KPI)s over live cellular networks based on the combination of Recurrent Neural Networks (RNN), and Convolutional Neural Networks (CNN), as Recurrent Convolutional Neural Networks (R-CNN). CNN models the spatial correlations and information, whereas, RNN models the temporal correlations and information. Hence, adopting R-CNN provides us with spatial-temporal analysis. In this paper, the studied cellular network consists of 2G, 3G, 4G, and 4.5G technologies, and the KPIs include Voice and data traffic of the cells. The data and voice traffics are extremely important for the owner of wireless networks, because it is directly related to the revenue, and quality of service that users experience. These traffic changes happen due to a couple of reasons: the subscriber behavior changes due to especial events, making neighbor sites on-air or down, or by shifting the traffic to the other technologies, e.g. shifting the traffic from 3G to 4G. Traditionally, in order to keep the network stable, the traffic should be observed layer by layer during each interval to detect major changes in KPIs, in large scale telecommunication networks it will be too time-consuming with the low accuracy of anomaly detection. However, the proposed algorithm is capable of detecting the abnormal KPIs for each element of the network in a time-efficient and accurate manner. It observes the traffic layer trends, and classifies them into 8 traffic categories: Normal, Suddenly Increasing, Gradually Increasing, Suddenly Decreasing, Gradually Decreasing, Faulty Site, New Site, and Down Site. This classification task enables the vendors and operators to detect anomalies in their live networks in order to keep the KPIs in normal trend. The algorithm is trained and tested on the real dataset over a cellular network with more than 25000 thousand.

## I. INTRODUCTION

Over the last few years, the tendency toward utilization of internet-based services such as the Internet of Things(IoT), internet-based mobile application, and the user's demands to have access to the internet every time and everywhere leads to make the cellular networks larger in terms of covered areas and number of subscribers. Therefore, the maintenance and monitoring of the performance of these networks adopting traditional approaches require vast resources, on the other hand, each degradation or failure on cellular network in each layer (cell, site, or cluster), affects the subscriber's experience, and it may result in an irrecoverable event. For all Mobile Operators, the changes in traffic or payload amounts are of utmost importance, because it shows an event, or it comes up with the revenue, on the other hand, the increment or reduction of the traffic, directly affects networks' profits or losses. Along with traffic behavior detection, traffic prediction is also an important research problem, some regular events or subscriber behavior can be predicted thus the required resources or strategies can be planned. Besides, the precise estimation of traffic volume is useful for congestion control [1], network routing and resource allocation. Therefore, detection and respond of anomalies of cellular network performance in real-time is critical for network administrators, so, in order to improve the subscriber service quality or maintaining network dependability. On the other hand, one of the main tasks of network administrators is to pinpoint any traffic/payload abnormality, which refers to unexpected patterns occur at a single time instant or over a long time period.

Some surveys were published which evaluates the utilization of machine learning approaches for performance analysis of cellular mobile networks. In [2] the author present CELL- PAD, it is a performance anomaly detection platform for time- series KPI analysis. It realizes simple statistical modeling and regression as a machine-learning-based algorithm for anomaly detection. [3] is a survey of Machine

Learning on SON (Self Organizing Cellular Networks). It provides the comparison between performances of the SON machine learning algorithms, besides it proposes the best performance ML algorithm for each SON function. In [4], the authors propose a scheme of Robust Statistical Traffic Classification (RTC) by combining unsupervised and supervised machine learning techniques. This traffic classification scheme is proposed to tackle the problem of zero-day applications. It focuses on application, port and IP traffic. The two machine learning traffic classification methods are compared in [5], the supervised Support Vector Machine (SVM) and unsupervised K- means clustering. This paper also considers the application, port and IP payload. The [6] presents a used case AI (Artificial Intelligence) application, it tells that as 80 percent of power consumption is used in RAN in general and BS (Base Station) as particular, on the other hand, when BS takes the lowest traffic in the early morning (from 6 AM to 8 AM), the BS with AI switching policy costs 55 percent of the energy which would be consumed if it employs no energy saving scheme. By this approach the traffic has been predicted then the BSs incur minimal energy consumption. The [7] presents a comprehensive survey of the crossovers between the two areas, the ML and cellular networking. In [8], reinforcement learning (RL) is used to learn how to improve the downlink SINR through the exploration and exploitation of various alarm corrective actions. The [9], proposed a method to improve the performance of the downlink coordinated multipoint (DL CoMP) in heterogeneous fifth generation new radio (NR) networks. The function is applied through applying online deep learning to physical layer measurements in a realistic NR FDD network. The application of machine learning techniques for performance prediction in wireless networks is discussed in [10]. Applying the method to the management of wireless mobile networks has the potential to significantly reduce operational costs while simultaneously improving user experience. Deep learning is a promising solution for the task of network performance analysis, because of its ability to distinguish the inherent relationships between the inputs and outputs without human involvement.

This paper presents a scalable data-driven platform based on deep Recurrent Convolutional Neural Network (R-CNN). This model can be used to recognize all the traffic abnormal behaviors. For the classification task, 8 different classes are assumed for this model, suddenly increasing which refers to a profile in which the KPI has been decreased noticeably in a short period of time, gradually increasing which is related to the profile in which the KPI has been degraded gradually in a longer period of time, which illustrates a problem in the network that lasts for a long time. Suddenly increasing and gradually increasing KPI's importance makes sense in the situations where, for example, drop call rate is the main KPI. In this situation, the augmentation in the amount of KPI will penalize the network performance. There will be a normal class for the KPIs related to the cells which perform in their normall trends. The data for this paper has been labeled manually for more than 2500 cells.

The rest of the paper is organized as follows. In Section II, the system model and the optimization problem statement are presented. Novel multi-agent DRL scheme and the framework of mapping the environment onto the 2D RGB virtual image are discussed in Section III. Simulation results are presented in Section IV, and the paper is concluded in Section V.

## II. DEEP LEARNING OVERVIEW

Deep learning is a type of machine learning that can be categorized into supervised, self-supervised, semi-supervised, and reinforcement learning types. "Fig. 1" shows a simple neural network and a deep learning neural network, it illustrates that deep learning is a type of neural network with more than one hidden layer.

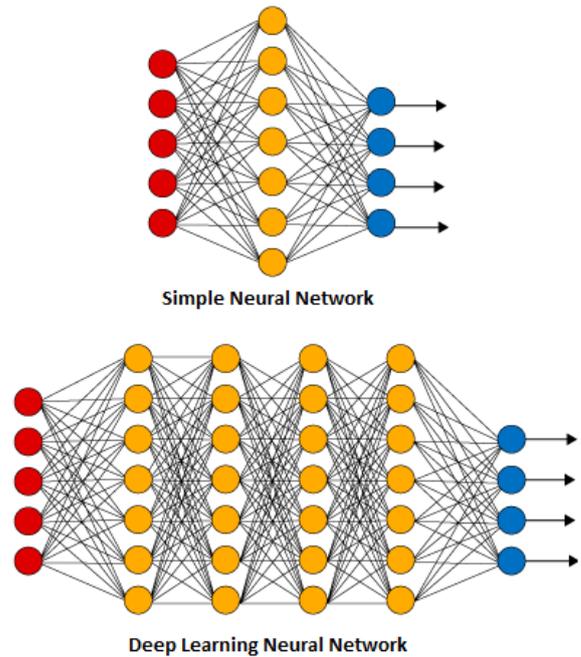

Fig. 1. Example of a shallow and deep multi-layer perceptron

A simplified of artificial neuron is a function $f_j(x)$ on input $(x_1, \ldots, x_n)$, which are weighted by a vector of $\Theta_j = (\Theta_{1,j}, \ldots, \Theta_{n,j})$ and completed by a bias vector $b_j$, and associated to an activation function $\varphi$, namely

$$f_j(x) = \varphi(<\Theta_j, x> + b_j)$$

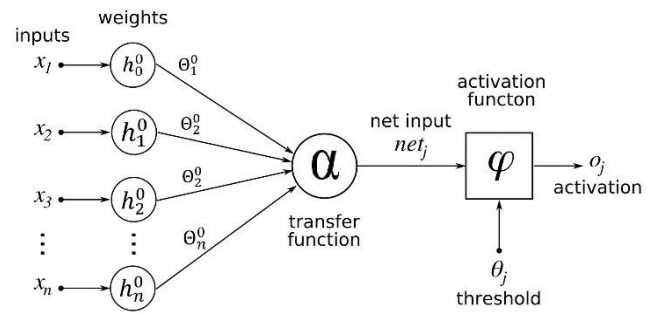

Fig. 2. Deep neural network architecture with activation functions

In the activation part several functions can be used.

- The step function
- The logistic or sigmoid function
- The tanh (hyperbolic tangent)
- The ReLU (Rectified Linear Unit)

The "Fig.3" illustrates these activation functions state.

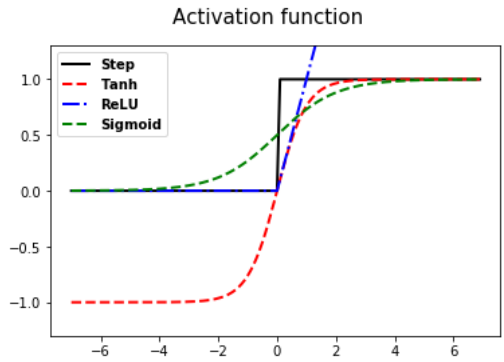

Fig. 3 different types of the activation function

A multi-layer perceptron (MLP) network is a structure includes several layers of neurons, in which the output of a neuron of a layer is the input of a neuron of the next layer. Finding the optimum weights by training is the aim of this kind of network.

The most important parts of MLP are listed as:

- Loss functions
- Backpropagation algorithm
- Regularization techniques

*A. A. Loss functions [9]:*

Depending on the problem type, different loss functions are being used. As this application is a type of classification algorithm, the Cross-Entropy is used as its loss function. The cross-entropy loss functions output is a probability value between 0 and 1, which is defined as (II)

$$J(\Theta) = \sum_{t=0}^{T_{mb}-1} J_{mb}(\Theta)$$

$T_{mb}$ is the number of training examples in a batch.
$t \in [0, T_{mb} - 1]$ is the batch training instance index.
$\Theta$ is the weights matrices.
N is the number of layers without input.
$F_v$ is the number of neurons in the $v$'th layer.
$y_f^{(t)}$ where $f \in [0, F_N - 1]$ is the output variables (to be predicted)
$\hat{y}_f^{(t)}$ where $f \in [0, F_N - 1]$ is the output of the network

*B. Backpropagation algorithm*

To minimize the criterion expected loss, a stochastic gradient descent algorithm (SGD) is used. And to compute the gradient, the backpropagation (BP) algorithm is considered.
BP for output layers:

$$\delta_{fc}^{(t)(N-1)} = \frac{1}{T_{mb}} (h_f^{(t)(N)} - \delta_{y^{(t)}}^f)$$

BP for hidden layers:

$$\delta_f^{(t)(v)} = g'(\alpha_f^{(t)(v)}) \sum_{t'=0}^{T_{mb}-1} \sum_{f'=0}^{F_{v+1}-1} (\Theta_f^{(v+1)f'} J_f^{(tt')(v)} \delta_{f'}^{(t)(v+1)})$$

*C. Regularization techniques*

One of the difficulties when coping with deep learning techniques is to get a deep neural network to train effectively. For it several regularization techniques have been introduced. Some of them have been listed as follows:

- L2 regularization
- L1 regularization
- Clipping
- Dropout
- Batch Normalization

In this paper, Dropout and Bach Normalization techniques are used to get better accuracy in the training and test process. The dropout technique enables us to ignore some neurons in hidden layers during the training process to avoid the overfit problems. The batch normalized technique can normalize each layer and helps us to train the normalization weights. Generally, the input data is fed into the model at the input layer of the neural network. For the feed-forward network, each layer's output is obtained based on the input data and initializing random weights, then the error is calculated adopting proper loss function. Next, calculating the derivatives, the back-propagation is being implemented to update the weights of the whole network, which basically is based on loss minimization. During this process of, back-propagation and updating the weights, the optimum weights for the neural network will be achieved.

### III. SUPERVISED LEARNING

Supervised learning refers to a category of learning algo-rithms, in which both the labels and the data are in hand. Supervised learning algorithm has major applications in artificial intelligence, medical image analysis, pattern recognition, and wireless networks. However, in some applications due to having no sufficient labeled data, researchers seek to use other methods like self-supervised learning, semi-supervised learning and reinforcement learning. Self-supervised learning is related to a category of learning algorithms in which there are no labeled data. The other category is related to semi- supervised learning, in which there are a few labeled data, but the number of labeled data in comparison to labeled data is little. Also, reinforcement learning (RL) relates to a class of algorithms based on actions of an agent in an environment, and there are neither data nor labels beforehand. The neural net in RL is being trained based on action-reward pairs. The agent takes action and receives a reward/punishment from the environment. In this section, we consider the different parts of the supervised learning algorithm. Two categories of neural networks which are mostly being used in supervised manner are Convolutional Neural Network (CNN), and Recurrent Neural Network (RNN). CNN models the spatial correlations and information, however, RNN models the temporal correlations and information. Hence, adopting recurrent convolutional neural networks provides us with spatial-temporal analysis.

*A. Convolutional Neural Network (CNN)*

CNN is planned by several convolutional layers and pooling which is followed by one or more fully connected layers for the classification tasks. We should have labeled data for the training and test process to train the network and measure the test accuracy in a supervised manner. The Convolutional Layer consists of a set of filters which in convoluted with the data matrices. The weights of the mentioned filters are being learned to detect the patterns or specific features in the original data. These filters are captured or convolved across the input file, and new data is computed for the activation map. The pooling layer tends to reduce the amount of computation, parameters and controlling the over-fitting by progressively reducing the spatial size of the network. In "Fig. 4" a CNN network has been illustrated [7].

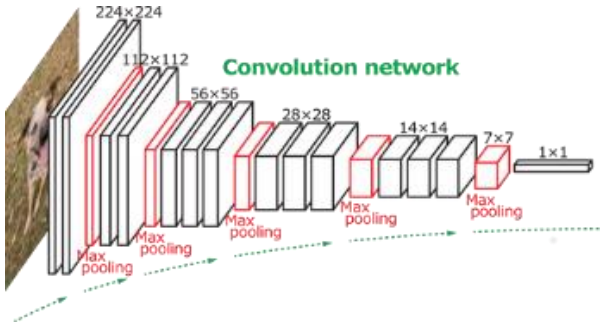

Fig.4 Convolutional Neural Network

## B. Recurrent Neural Network (RNN)

The most important characteristic of RNNs is using the memory to process sequences of inputs. It means, unlike the CNN architecture, where there is no memory to capture the temporal correlation, RNNs have the capabilities to capture the sequential correlations and information. There are 3 types of recurrent neural network: the simple (RNN), long short term memory unit (LSTM) and gated recurrent unit (GRU). The other advantage with RNNs is that the input size of the data, i.e., there is no limitation for the input and output size and structure. One of the most prestigious RNN systems is Long-Short Term Memory (LSTM) which has been used in this work to model the network KPI's temporal behaviors.

In LSTM networks, there is a cell memory. After training the neural net, the network can select the information to save to pass through the networks, and which part of the information can be ignored. Hence, there is a forget gate and memory cell [11, 12].

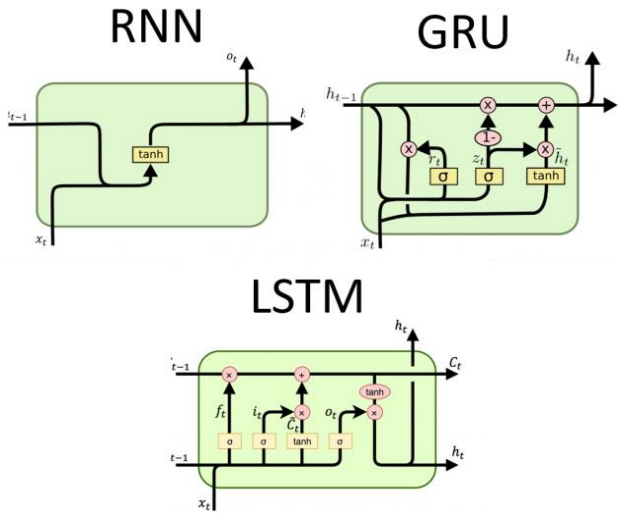

Fig 5. Recurrent Neural Network types.

In LSTM the follows notations have been defined:
$h_t$, $C_t$ : Hidden layer vectors,
$x_t$ : Output vector,
$b_f, b_i, b_c, b_o$ : Bias vector,
$W_f, W_i, W_c, W_o$ : Parameter matrices, and
σ, tanh : Activation functions.

## IV. MTN IRAN-CELL CHARACTERISTICS

This tool platform is a suitable platform for being used in Mobile Operators like MTN Iran-cell, which is officially titled Iran's greatest data operator; Iran-cell is pioneering in equipping its subscribers with the fastest and most advanced data network in the country. Besides, Iran-cell owns the largest 2G-3G-4G-4.5G RAN and modernized core mobile network, and fixed wireless TD-LTE internet services in the Middle East. It connects over 45 million people in Iran via 260,000 cell carriers in different technologies.

*1) Performance and Optimization Challenges*: As the Mobile networks advance to such large scales, the operation, optimization, configuration, and management teams will be in dire need to have intelligent tools to handle the new and diverse problems. On the other hand, such huge mobile operators these types of networks will also collect gigantic amounts of data in order to monitor, maintain and improve network stability, performance to provide better services by performing optimization actions to satisfy subscribers' expectation. In traditional approaches, some applications are used just to monitor main KPIs or revenue streams of the operator to ensure operator obligations are met and investment is protected. Based on our experiences, the particular KPI data and voice traffic - can be observed and monitored in these applications. The patterns of traffic should be extracted as data files, and do the pre-processing on them to find the behavior. This is a time-consuming process, and at the same time it is not efficient for network improvement.

*2) Application Specification:* The input to our proposed platform is a set of RAN KPIs as time-series data. Main Tools characteristics are as follows:

A) The proposed plat-from is scalable since it is capable of detecting anomalies regardless of the type of KPI. It means, it is possible to input any type of KPI to the trained neural network to capture the abnormality.
B) The technology which is used can be different network to network, and the model can capture the abnormalities, so it can be used for all kinds of technologies such as 2G, 3G, 4G, 4.5G, and 5G.
C) The network layer in which the task of anomaly detection is performed is not important as it can detect and classify patter regardless of the geographical layer in which statistics are aggregated.
D) The output of the model is not dependent on the duration of the time interval, in which the KPI has been measured. Thus, the time interval can be hourly/daily/weekly/monthly.
E) The size of the input is not important, for example, the abnormal behavior of one KPI with 100 sample times in MTN Irancell for 260,000 cells can be categorized into eight different classes which are already mentioned, takes less than one minute on Laptop T470P with 8GB RAM, CPU Core i7, although it works on GPU.

## V. SYSTEM MODEL

### A. Data Set

The dataset has been gathered from MTN Iran-cell live cellular network with more than 260000 cells. The data has been labeled for 2000 cells. As illustrated in "Fig. 6", the data has been labeled manually into eight categories. The time interval for the data of each cell varies from 6 months to one week in order to avoid overfitting.

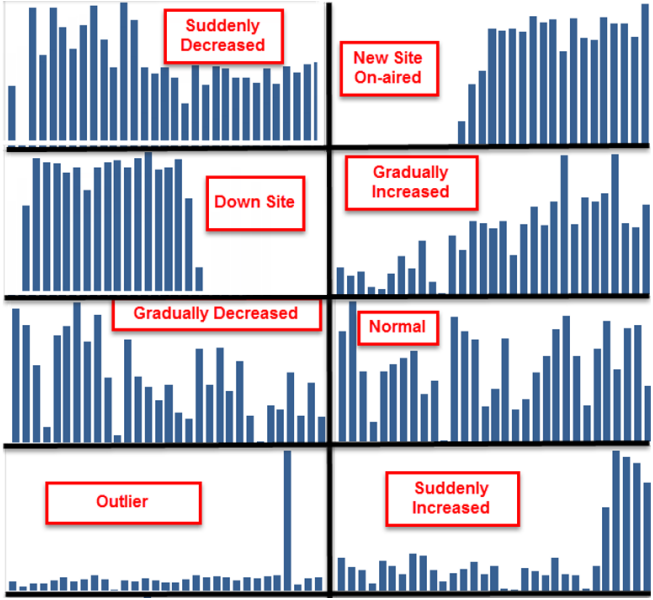

Fig. 6. Different cells with corresponding KPIs that labeled manually into one of the 8 classes.

### B. Network Architecture

The proposed deep neural network for anomaly detection in the live cellular network consists of two parts: convolutional neural nets and recurrent neural net. RNN parts model the temporal information in KPIs data, whereas, CNN part models the spatial information in KPIs patterns. As illustrated in "Fig. 7", the first part is a CNN, followed by an LSTM network. Next, there is a classifier part to classify the input KPI signal into one of the eight possible classes.

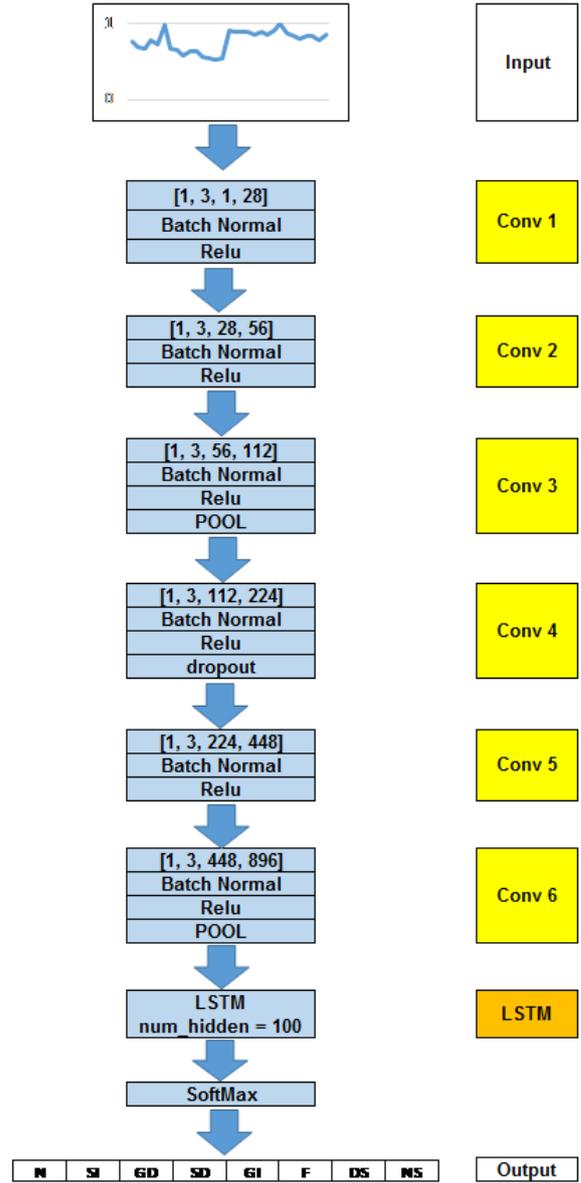

Fig. 7. Proposed network structure: R-CNN

### C. Results

In this paper, we provide two different approaches, for the task of anomaly detection based on network KPIs over live cellular networks, first we utilize the CNN network without any RNN cells, and the proposed approach in which both the RNN and CNN have been used simultaneously. The table shows the statistics on accuracy.

TABLE I. TABLE TYPE STYLES

| Methods | Average | |
| --- | --- | --- |
| | Accuracy | Loss |
| CNN | 0.942 | 27452 |
| RCNN | 0.932 | 0.212 |

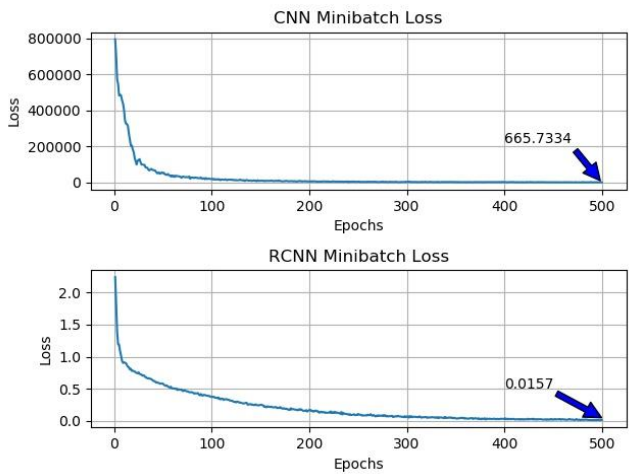

Fig. 8. Proposed network structure: R-CNN

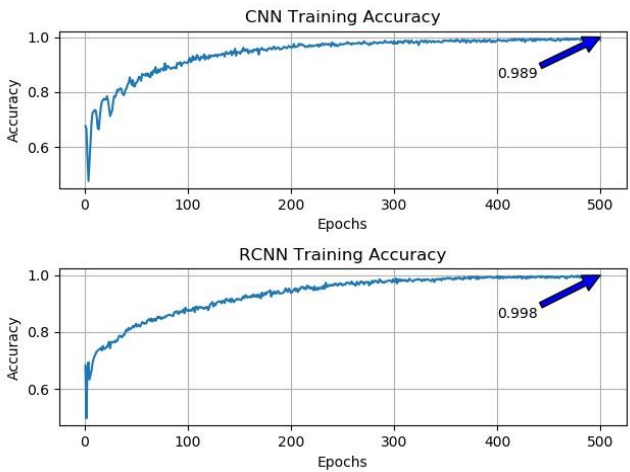

Fig. 9. Proposed network structure: R-CN

## VI. CONCLUSION

In this paper, a Recurrent Convolutional Neural Network is proposed in order to detect anomaly KPIs in a live cellular network, with more than 260000 cells. The RCNN architecture modeled both the spatial and temporal data in KPI trends. It is showed that the accuracy for this task is almost 98 percent. It provides the operators with a facility in order to track the pattern in their data and voice traffic, in order to detect abnormal changes in proper time. In comparison to only using CNN architecture, RCNN outperforms in both the accuracy and recall metrics. This method is scalable, and can be run in clusters in a distributed manner.